# Self-Aligned Ballistic Molecular Transistors and Electrically Parallel Nanotube Arrays

Ali Javey[1], Jing Guo[2], Damon B. Farmer[3], Qian Wang[1], Erhan Yenilmez[1], Roy G. Gordon[4], Mark Lundstrom[2], Hongjie Dai[1*]

[1] Department of Chemistry and Laboratory for Advanced Materials, Stanford University, Stanford, CA 94305, USA

[2] School of Electrical and Computer Engineering, Purdue University, West Lafayette, IN 47907, USA

[3] Division of Engineering and Applied Sciences, Harvard University, Cambridge, MA 02138, USA

[4] Department of Chemistry and Chemical Biology, Harvard University, Cambridge, MA 02138, USA

## Abstract

Carbon nanotube field-effect transistors with structures and properties near the scaling limit with short (down to 50 nm) channels, self aligned geometries, palladium electrodes with low contact resistance and high-κ dielectric gate insulators are realized. Electrical transport in these miniature transistors is near ballistic up to high biases at both room and low temperatures. Atomic layer deposited (ALD) high-κ films interact with nanotube sidewalls via van der Waals interactions without causing weak localization at 4 K. New fundamental understanding of ballistic transport, optical phonon scattering and potential interfacial scattering mechanisms in nanotubes are obtained. Also, parallel arrays of such molecular transistors are enabled to deliver macroscopic currents – an important milestone for future circuit applications.

* Correspondence to hdai@stanford.edu

Single-walled carbon nanotubes (SWNT) have provided researchers with excellent model systems for elucidating fundamental properties of quasi one-dimensional (1D) materials, and triggered interesting questions such as what the ultimate 1D electronics (such as field effect transistors, FETs) might be.[1-8] Ballistic transport, a desired property for high performance electronics, has been demonstrated for SWNTs in the low bias regime[9-13] but remains unclear whether it can be achieved in high-bias operations of nanotube transistors, especially in real devices for which potentially damaging processes such as dielectric deposition become indispensable.

It has been suggested that high-κ dielectrics might be essential to future transistors due to high gate capacitance, low leakage currents and power dissipation.[14] However, a fundamental problem for conventional semiconductors is the degradations of electrical properties due to carrier scattering mechanisms introduced at the high κ film-semiconductor interface.[15] For example, silicon MOSFETs with deposited high-κ dielectrics consistently display drastically inferior properties compared to those with thermally grown $SiO_2$ gate insulators.[16] In the current work, we show that semiconducting carbon nanotubes represent the first exception in affording nearly ballistic transistors with high-κ dielectrics, opening the door to ultra-fast electronics since both ballistic transport and high-κ dielectrics facilitate high ON-current that is directly proportional to the speed of a transistor.

We also demonstrate ultra-short molecular transistors with self-aligned S, D and G structures that once represented an important milestone[17] for CMOS technology. Our devices consist of L~50 nm long SWNTs between palladium S and D contacts, 8 nm thick $HfO_2$ high-κ (κ~15) gate insulator formed on top of SWNTs by atomic layer deposition (ALD) at 90 °C[18] and top Al gate electrodes (Fig. 1a). Self-alignment means that the edges of the S, D and G electrodes are precisely and automatically positioned such that no overlapping or significant gaps exist between them (Fig. 1a). This was made possible by two key steps. The first was the development of ALD at 90 °C[18] allowing for deposition of high-κ films on substrates patterned with a polymer-resist PMMA.[19] High-κ dielectric 'lines' (~8 nm thick; width ~ 50 nm defines channel length L), topped by Al gate metal (~50 nm thick) were first formed (as a gate-stack) by the liftoff method[19] to cover SWNTs (Fig. 1a). The second key step takes advantage of native $Al_2O_3$ (4-8 nm thick) on the Al metal gate.[20] Pd metal (thickness ~ 7nm) deposited in the region became divided by the high-κ/Al/$Al_2O_3$ gate stack, forming the S and D Pd[7] electrodes perfectly aligned on the two sides of the gate stack (Fig. 1a&b). The insulating $Al_2O_3$ film on the Al gate and the directional deposition of thin Pd ensured electrical insulation between G, S and D, but a series resistance of ~ 1.7 kΩ existed for each of the S/D electrodes due to the thin Pd (7 nm, width ~8 μm, length ~200 μm). Our method is capable of fabricating self-aligned p-type SWNT FETs with arbitrary channel length. The self-aligned structure minimizes parasitic capacitances and will be indispensable for high-speed operations.

Our miniaturized self-aligned SWNT FETs with high-κ $HfO_2$ exhibit high peak transconductance $(dI_{DS}/dV_G)_{max}$ ~ 30 μS per tube, maximum linear ON-state conductance of ~$0.5 \times 4e^2/h$ and saturation current up to ~25 μA (Fig.1c&d). The saturation current is the highest reached for any SWNT-FETs, notably under the lowest bias of $V_{DS}$~0.4 V. The ON and OFF ratio for the SWNT (diameter $d$~1.7 nm) is $I_{ON}/I_{OFF} > 10^3$ at $V_{DS}$=0.3 V with a subthreshold swing of ~110 mV/decade (Fig.1c&d). Despite the series resistance, these characteristics collectively represent the best for nanotube FETs.

We have theoretically modeled the devices used in experiments by solving Poisson's equation in 3D for electrostatics and by assuming fully ballistic transport[21] with zero Schottky barrier (SB) for holes at the Pd contacts.[7] With a series resistance of 1.7 kΩ per S/D contact (due to thin Pd) included in simulation (Fig. 1c&d symbols), the result matches the experiment (Fig. 1c&d, solid lines) well. The experimental currents at the high-bias end are slightly lower than theory (Fig. 1d), attributed to slight inelastic optical phonon scattering.[11-13] Even without correction for the series resistance, the maximum theoretical (truly ballistic) current is only 20% higher than the measured current. These comparisons suggest that the experimental FET delivers DC currents close to the ballistic limit, consistent with the SWNT length L~50 nm significantly below the mean free path (mfp) of $L_{ap}$~300 nm and $L_d$ ~ 1μm for elastic acoustic phonon (at 300 K) and defect scattering, respectively.[12] The near-ballistic current appears high given that the SWNT length L~50 nm is about three times the mfp of $L_{op}$~15 nm for optical phonon scattering expected at high biases.[11-13] Our simulations show however, that when carriers lose energy by optical phonon emission, they are unlikely to return to the source due to the reduced energy and the potential profile in the tube (J. Guo and M. Lundstrom, in preparation). Inelastic scattering events therefore, have small effects on the DC current in semiconducting SWNTs several times longer than the inelastic scattering mfp. The near-ballistic transistors suggest that deposition of high-κ dielectric films does not harm the room temperature electrical characteristics of SWNTs. When cooled, the p-channel conductance of our SWNT FETs exhibited no significant temperature dependence (Fig.2a) with the appearance of Fabry-Perot[9] type of resonance at 4 K (Fig. 2b), signaling ballistic transport in the ohmically contacted p-channel (SB to p-channel ~0 with Pd)[7] at low temperatures. No conductance lowering or sharp random fluctuations due to weak localization[10,22] were observed. The n-channel also exhibited no significant temperature dependence until 100 K, below which Coulomb oscillations (Fig. 2b) due to single electron charging were observed, corresponding to a quantum dot confined by the Schottky barriers (SB) at the Pd contacts to the n-channel of the SWNT (barrier height ~ band gap $E_g$ ~ 0.5 eV). The periodic Coulomb oscillations (Fig. 2b inset) corresponded to a single dot, again indicating the lack of significant disordering in the nanotube due to deposited high-κ film.

The ballistic and phase coherent transport at low temperature is remarkable considering the deposited high-κ material on the nanotube sidewall and that 1D systems are susceptible to weak localization even for small amount of disorder.[22] Apparently, any perturbation caused by the ALD process does not adversely affect the electrical properties of SWNTs even at 4 K. This result sheds significant light on the formation of the dielectric/nanotube interface. The interaction between the deposited high-κ films and the SWNT sidewalls should be van der Waals (vdW) in nature without the formation of covalent bonds in any random or homogeneous fashion along the tube, at least at the L=50 nm scale. Deposition of high-κ films on a SWNT must be nucleated on the $SiO_2$ substrate surrounding the SWNT and then grown to cover the tube. It is in fact known that high-κ film deposition on $SiO_2$ starts with chemisorption between ALD precursor molecules and –OH groups on $SiO_2$ surface.[18] No uniform film can be grown on SWNTs without a supporting substrate, for instance, in the case of suspended nanotubes (Fig. 2c).

The lack of surface dangling bonds and chemical inertness mark a fundamental difference between carbon nanotubes and other semiconductors known. In Si devices, carrier mobility degradation is attributed to Coulomb scattering by charges residing in the high-κ/Si interface states, soft optical phonons in the high-κ film and surface roughness.[15] For nanotubes, the vdW interface with the high-κ film should afford negligible scattering by interface states and surface roughness. Scattering by soft phonons in the high-κ film also appears to be insignificant as signaled by the near ballistic transport in our SWNT-FETs. This could also be due to the weak non-covalent interface between the nanotube and high-κ film. We hope that our current work will stimulate theoretical investigations of scattering effects in nanotubes caused by phonons in high κ films.

We further developed a novel strategy to obtain multiple self-aligned ballistic FETs on a single nanotube by using inter-penetrating comb-like S, D and G electrodes (Fig. 3a&b). Multiple gate stacks were first patterned on a single SWNT with additional alternating 'connectors' between the gate stacks (Fig. 3a&b). Deposition of Pd in the region led to self-aligned Pd S/D electrodes, divided and isolated by the connector lines. The S (or D) electrodes for all the FETs on the same side of the connector lines were shorted together, effectively affording nanotube FETs in a parallel array. With 8 self-aligned ballistic FETs electrically connected in parallel, we were able to obtain a device capable of delivering over 150 μA (Fig. 3c&d). This illustrates the high reproducibility of individual self-aligned near-ballistic FETs, and more importantly, demonstrate for the first time that arrays of nanotube FETs can deliver macroscopic currents, which is a critical step towards practical circuit applications.

**Acknowledgements**

We thank Charis Quay and David Goldhaber-Gordon for use of low temperature measurement instruments. This work was supported by MARCO MSD Focus Center, Stanford INMP, DARPA Moletronics, SRC/AMD, DARPA MTO, a Packard Fellowship, NSF Network for Computational Nanotechnology, and an SRC Peter Verhofstadt Graduate Fellowship (A. J.).

**Figure Captions**

**Figure 1. Self-aligned near-ballistic SWNT FETs.** (a) Side-view schematic of a device. SWNTs were grown by chemical vapor deposition[23] on Si($p^{++}$)/SiO$_2$ substrates. ALD of HfO$_2$ used tetrakis(diethylamido)hafnium (Hf[NEt$_2$]$_4$) as precursor. For each of the 80 ALD cycles (~ 0.1 nm/cycle) used, the purge times were 350 s after the DI H$_2$O dose, and 150 s after the Hf[NEt$_2$]$_4$ dose. The deposition of Pd was by highly directional electron beam evaporation. For all of our measurements, the bottom gate (Si substrate) was grounded. (b) Scanning electron microscopy (SEM) image showing the top-view of a device. The nanotube appears faint under the thin Pd electrodes. (c) Current vs. top-gate voltage ($I_{DS}$-$V_G$) for a device with a L~50 nm and d~1.7 nm SWNT at different biases ($V_{DS}$). The devices were annealed in Ar at 175 °C for 5 min to obtain optimum Pd-SWNT contacts. PMMA passivation[24] was used for the measurements. The annealing and passivation treatment steps afforded up to 2-5 fold increase in the p-channel conductance of SWNT FETs. (d) $I_{DS}$-$V_{DS}$ characteristics of the same device. Solid lines are experimental data and symbols are ballistic quantum simulation in (c) (symbols correspond to $V_{DS}$=0.3V) and (d). Parameters used in simulations (J. Guo et al., to be published), band-gap $E_G$=0.5 eV (see Fig. 2), SB height for holes ~ 0, and geometrical parameters identical to experiments.

**Figure 2. Cooling of self-aligned and near-ballistic high-κ SWNT-FETs.** (a) Conductance (G) vs. $V_G$ of a L~50 nm and d~1.7 nm SWNT device recorded at different temperatures. Inset: resistance $R_{max}$ at the lowest conductance points in the G-$V_G$ curves at various temperatures (T) fitted to ln $R_{max}$ ~ $E_G$/2$k_B$T, giving rise to a band gap of $E_G \sim 0.5$ eV for the d~1.7 nm SWNT. (b) G-$V_G$ of the same device at 4 K. The gate efficiency was $\alpha = E_g/\Delta V_G(\text{gap}) = 0.6$ estimated from the band gap of the tube $E_g = 0.5$ eV and gap-width $\Delta V_G(\text{gap})$ in the $I_{DS}$ vs. $V_G$ curve in (a). The gate capacitance was $C_G = e/\Delta V_G(\text{CB}) \sim 4.3$ aF from the Coulomb oscillation period $\Delta V_G(\text{CB}) = 37$ mV from the inset in (b). The charging energy of the L~50 nm SWNT was $E_c = e^2/C_\Sigma = e^2/(C_G/\alpha) \sim 22$ meV. (c) A transmission electron micrograph (TEM) for a suspended SWNT (across slits in a nitride membrane) treated by the HfO$_2$ ALD process. The data shows that pristine nanotubes do not react with the precursors under ALD conditions to form uniform dielectric coating. An occasional defect site on the nanotube is likely to be responsible for the nucleation and growth the dielectric sphere seen.

**Figure 3.** An array of self-aligned and near ballistic SWNT-FETs connected in parallel. (a) & (b) SEM images of an array of FETs based on a single nanotube. (c) Transfer characteristics of a device at various S/D bias voltages. The device consists of a single nanotube crossing 8 gate lines, resulting in an array of 8 FETs. (d) Output characteristics of the same device showing over 150 µA of ON state current. The device was passivated with PMMA.

**References**

(1)     Dresselhaus, M. S.; Dresselhaus, G.; Avouris, P., Eds. *Carbon Nanotubes*; Springer: Berlin, 2001; Vol. 80.

(2)     Dekker, C.  *Phys. Today* **1999**, *52*, 22-28.

(3)     McEuen, P. L.; Fuhrer, M. S.; Park, H. K. *IEEE Trans. Nanotechnology* **2003**, *1*, 78-85.

(4)     Heinze, S.; Tersoff, J.; Martel, R.; Derycke, V.; Appenzeller, J.; Avouris, P. *Phys. Rev. Lett.* **2002**, *89*, 6801.

(5)     Wind, S.; Appenzeller, J.; Martel, R.; Derycke, V.; P, P. A. *Appl. Phys. Lett.* **2002**, *80*, 3817-3819.

(6)     Javey, A.; Kim, H.; Brink, M.; Wang, Q.; Ural, A.; Guo, J.; McIntyre, P.; McEuen, P.; Lundstrom, M.; Dai, H. *Nature Materials* **2002**, *1*, 241 - 246.

(7)     Javey, A.; Guo, J.; Wang, Q.; Lundstrom, M.; Dai, H. J. *Nature* **2003**, *424*, 654-657.

(8)     Javey, A.; Guo, J.; Farmer, D. B.; Wang, Q.; Wang, D. W.; Gordon, R. G.; Lundstrom, M.; Dai, H. J. *Nano Lett.* **2004**, *4*, 447-450.

(9)     Liang, W.; Bockrath, M.; Bozovic, D.; Hafner, J.; Tinkham, M.; Park, H. *Nature* **2001**, *411*, 665-669.

(10)    Kong, J.; Yenilmez, E.; Tombler, T. W.; Kim, W.; Liu, L.; Jayanthi, C. S.; Wu, S. Y.; Laughlin, R. B.; Dai, H. *Phys. Rev. Lett.* **2001**, *87*, 106801.

(11)    Yao, Z.; Kane, C. L.; Dekker, C. *Phys. Rev. Lett.* **2000**, *84*, 2941-2944.

(12)    Javey, A.; Guo, J.; Paulsson, M.; Wang, Q.; Mann, D.; Lundstrom, M.; Dai, H. *Phys. Rev. Lett.* **2004**, *92*, 106804.

(13) Park, J.-Y.; Rosenblatt, S.; Yaish, Y.; Sazonova, V.; Üstünel, H.; Braig, S.; Arias, T. A.; Brouwer, P. W.; McEuen, P. L. *Nano Lett.* **2004**, *4*, 517.

(14) *MRS Bull.* **March 2002**, *27*.

(15) Ren, Z.; Fischetti, M. V.; Gusev, E. P.; Cartier, E. A.; Chudzik, M. *IEEE IEDM* **2003**, p.33.32.31-34.

(16) Lochtefeld, A.; Antoniadis, D. A. *IEEE Elec. Dev. Lett.* **2001**, *22*, 95-97.

(17) Fair, R. B. *Proc. IEEE.* **1998**, *86*, 111-137.

(18) Hausmann, D. M.; Kim, E.; Becker, J.; Gordon, R. G. *Chem. Mater.* **2003**, *14*, 4350-4358.

(19) Biercuk, M. J.; Monsma, D. J.; Marcus, C. M.; Becker, J. S.; Gordon, R. G. *Appl. Phys. Lett.* **2003**, *83*, 2405.

(20) Bachtold, A.; Hadley, P.; Nakanishi, T.; Dekker, C. *Science* **2001**, *294*, 1317-1320.

(21) Datta, S. *Electronic Transport in Mesoscopic Systems*; University Press: Cambridge, 1995.

(22) Lee, P. A. *Rev. Mod. Phys.* **1985**, *57*, 287-337.

(23) Kong, J.; Soh, H.; Cassell, A.; Quate, C. F.; Dai, H. *Nature* **1998**, *395*, 878.

(24) Kim, W.; Javey, A.; Vermesh, O.; Wang, Q.; Li, Y.; Dai*, H. *Nano Lett.* **2003**, *3*, 193-198.

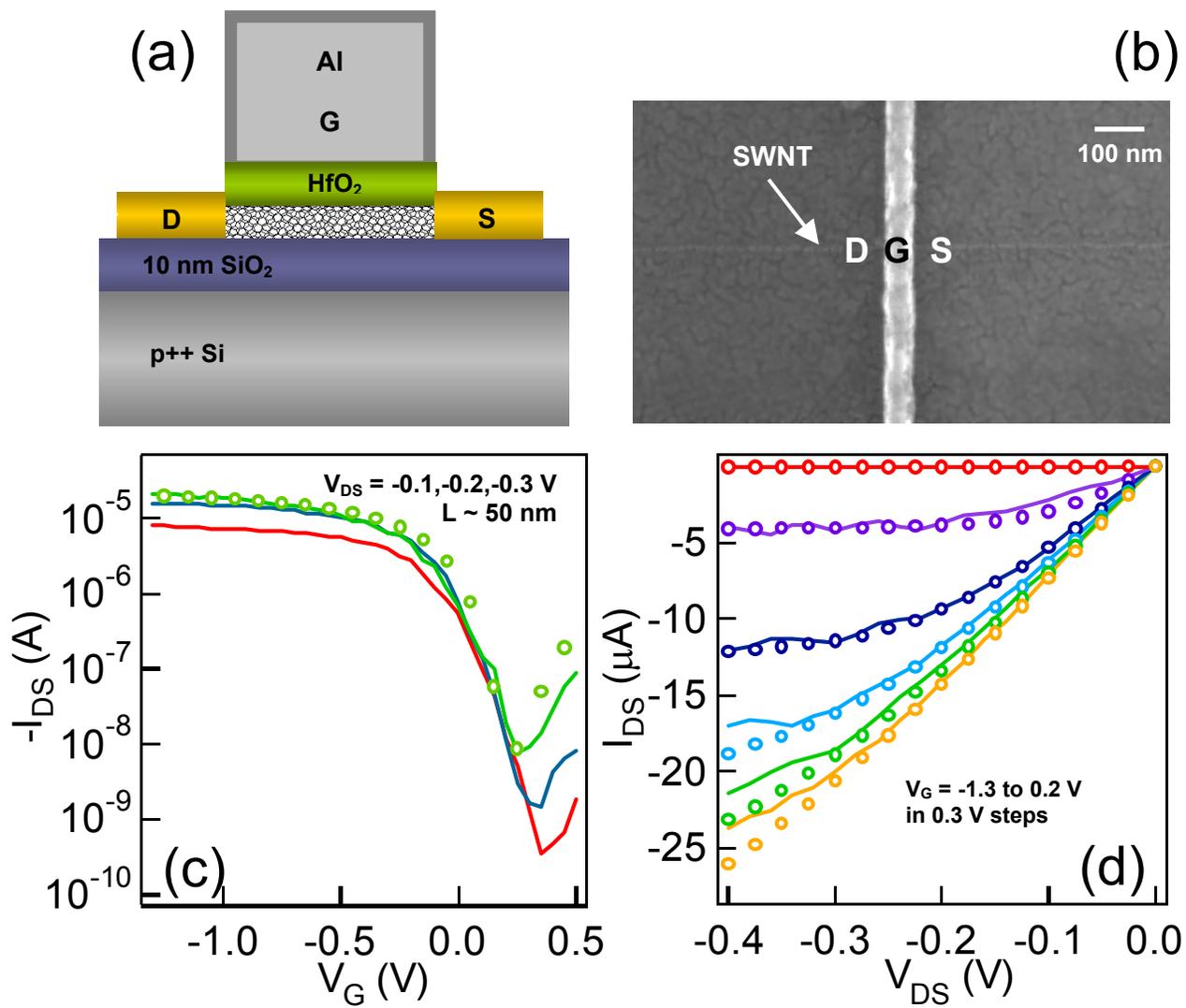

**Figure 1**

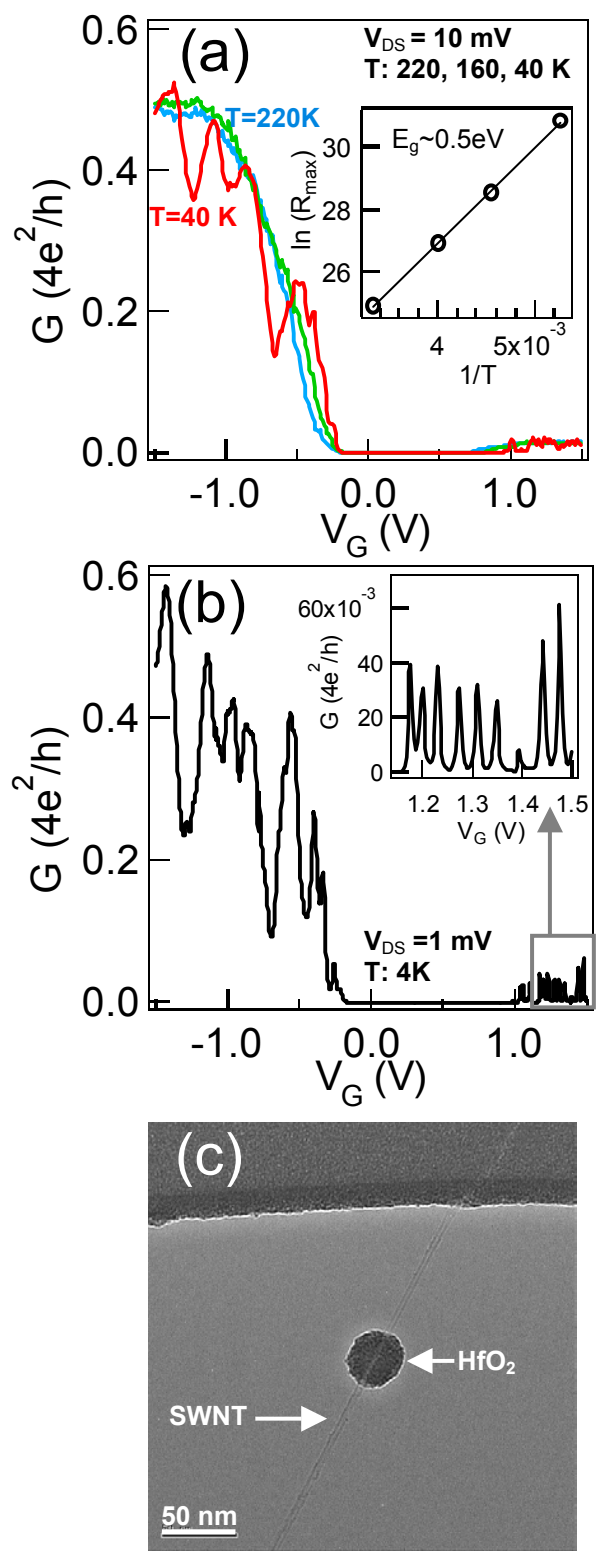

**Figure 2**

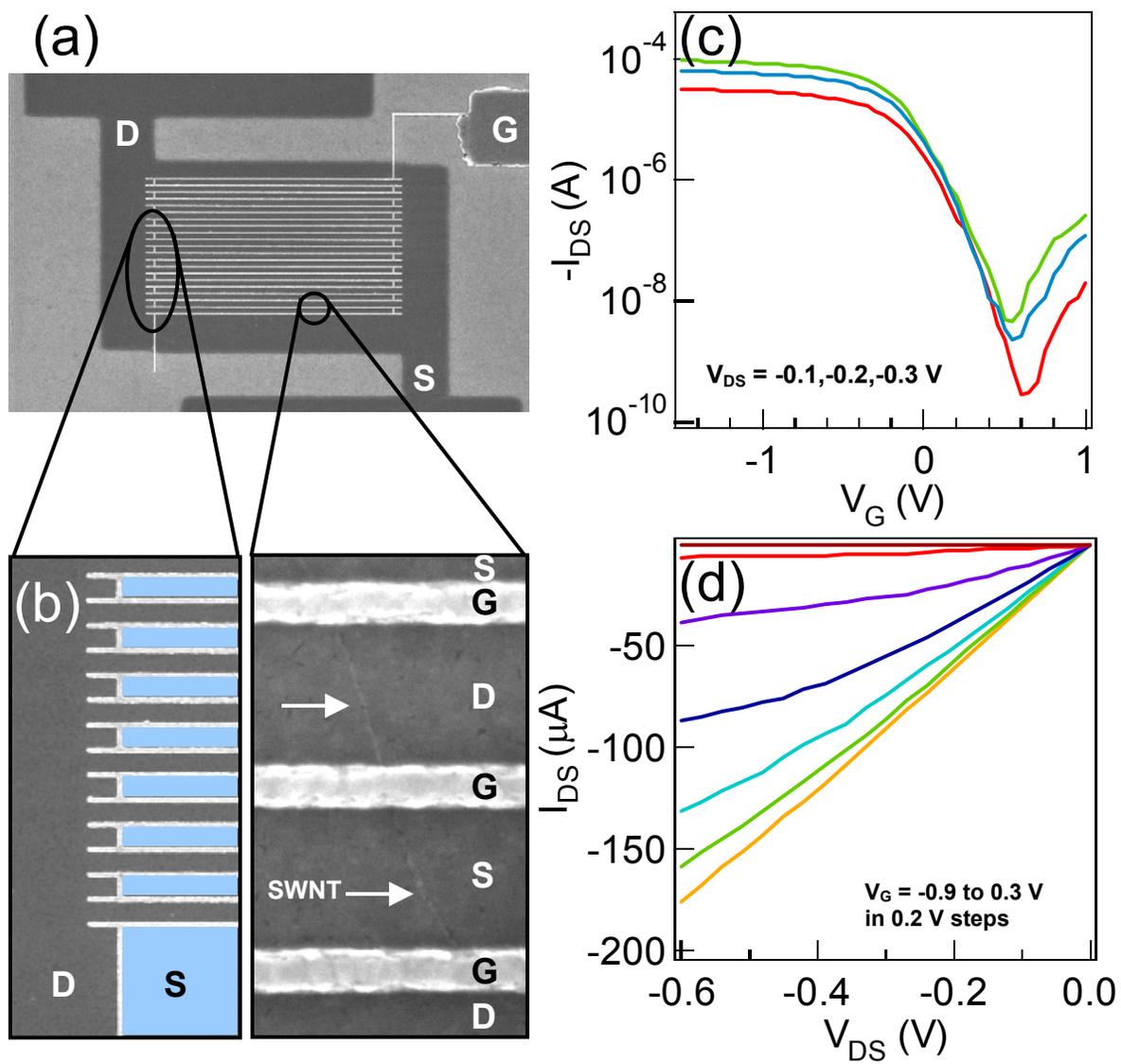

**Figure 3**

# Table of Contents

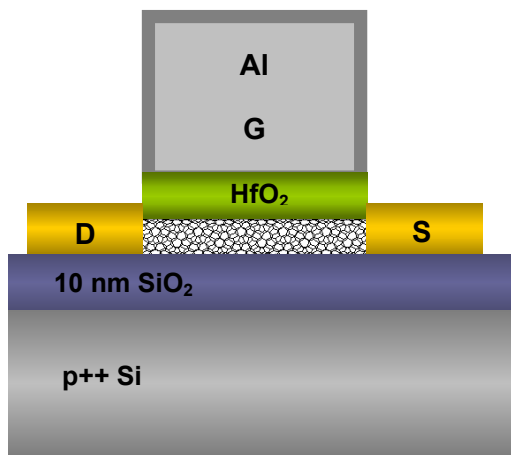 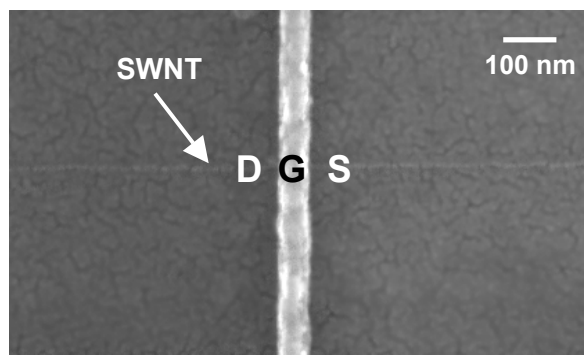